\documentclass[preprint,aps,prd,showpacs,superscriptaddress,nofootinbib]{revtex4}
\usepackage{graphicx,epsfig}
\usepackage{amsmath}
\usepackage{bm}
\usepackage{color}

\begin{document}


\title{\mbox{}\\[10pt]
Relativistic corrections to the semi-inclusive decay of
$\bm{\psi}$ and $\bm{\Upsilon}$}

\author{Hai-Ting Chen}
\affiliation{Key Laboratory of Frontiers in
Theoretical Physics, \\The Institute of Theoretical Physics, Chinese
Academy of Sciences, Beijing 100190, People's Republic of China}
\author{Yu-Qi Chen}
\affiliation{Key Laboratory of Frontiers in
Theoretical Physics, \\The Institute of Theoretical Physics, Chinese
Academy of Sciences, Beijing 100190, People's Republic of China}
\author{Wen-Long Sang}
\affiliation{Institute of High Energy Physics, Chinese Academy
of Sciences, \\Beijing 100049, People's Republic of China}
\affiliation{Department of Physics, Korea University, Seoul 136-701,
 Korea}




\begin{abstract}

In the framework of
the nonrelativistic quantum chromodynamics factorization formalism, we study the processes of $\psi(nS)$ and $\Upsilon(nS)$
decay into a
lepton pair or a charm pair associated with
two jets up to the next-to-leading order in velocity expansion.
We present the analytic expressions for the
differential decay rate to the invariant mass of the lepton pair or charm pair.
We find that
the ratio of the next-to-leading order short-distance coefficient to the leading order one
is in the range from $-5.5$ to $-12.4$. The relativistic corrections
are so large
that they modify the leading order prediction significantly.
Utilizing the analytic expressions, we also
investigate the relativistic corrections
in different kinematic regions
and their dependence on the
masses of the initial-state quarkonium
and the final-state fermion.
In addition, we study the
momentum distribution of $D^{*+}$ in the process
$\Upsilon(1S)\to c\bar{c}gg\to D^{*+}X$.

\end{abstract}

\maketitle
\newpage

\section{Introduction}
\label{introduction}

Heavy quarkonium decay phenomena have been extensively studied both in
theory and in experiment, from which one gains insight into both
the structure of the heavy quarkonium and quantum chromodynamics (QCD) interactions. The
predominant annihilation decay modes of the $S$-wave spin-triplet heavy
quarkonium are those hadronic decays, radiative decays, and leptonic
decays. With abundant data of  the $S$-wave spin-triplet heavy
quarkonium decays accumulated in experiments, higher order decay processes are
also interesting to investigate. Among them, two types of processes are
particularly interesting. One type is that the $S$-wave spin-triplet charmonium
and bottomonium semi-inclusive decay into a leptonic pair and light hadrons. The other
one is the $S$-wave spin-triplet bottomonium semi-inclusive
decays into a charm meson pair
and light hadrons.

In experiment, charm production via  $\Upsilon(1S)$ was studied
first by the ARGUS Collaboration~\cite{Albrecht:1992ap} and recently by
the \textsl{BABAR} Collaboration~\cite{:2009wm} as well as by the CLEO
Collaboration~\cite{Alex:2010}. \textsl{BABAR}'s results~\cite{:2009wm} provided
evidence for an excess of $D^{*\pm}$ production over the expected
rate from the virtual photon annihilation process $\Upsilon(1S)\to
\gamma^*\to c\bar{c}\to D^{*\pm}X$.  With a number of $\psi(nS)$
events accumulated at the Beijing Electron Positron Collider (BEPCII)~\cite{Asner:2008nq} and $\Upsilon(nS)$
events accumulated at B factories~\cite{Brambilla:2004wf}, the
$S$-wave spin-triplet charmonium and bottomonium semi-inclusive
decay into a lepton (charm) pair
and light hadrons are expected to be measured well.

In comparison with experimental data, it is necessary to
theoretically study those
processes precisely. The decay rate of these processes can
be analyzed in the framework of nonrelativistic QCD (NRQCD) factorization
formalism~\cite{Bodwin:1994jh}. According to it, the decay rates are
expressed as a sum of products of short-distance coefficients and
NRQCD matrix elements. The short-distance coefficients can be
expanded as perturbation series in coupling constant $\alpha_s$ at
the scale of the heavy quark mass. The long-distance matrix elements
can be expressed in a definite way with the typical relative
velocity $v$ of the heavy quark in the quarkonium state.

The decay rate of the semi-inclusive leptonic decay process $\psi(\Upsilon)\to
l^+l^-gg$ was first studied by J.~P.~Leveille and D.~M.~Scott in the
color-singlet model~\cite{Leveille-Scott:1980}. The polar and
azimuthal angular distributions of the lepton pair in this process
were also studied in Refs.~\cite{Korner:1982fn,Clavelli:1984ru}. The
semi-inclusive charm decay process $\Upsilon\to c\bar{c}gg$ was first
researched in Refs.~\cite{Fritzsch:1978ey,Parkhomenko:1994xi}, 
and the invariant mass
distribution of $c\bar{c}$ has been studied in
Ref.~\cite{Kang:2007uv}. The inclusive charm production in
$\Upsilon(nS)$ decay was calculated in Ref.~\cite{Chung:2008yf}.
Bigi and Nussinov have taken into account the contribution of
$\Upsilon\to c\bar{c}g$~\cite{Bigi:1978tj}. The exclusive double
charmonium production from $\Upsilon$ decay was calculated by
Jia~\cite{Jia:2007hy}. The authors of Ref.~\cite{Zhang:2008pr} also
considered the $\Upsilon$ decay to two charm jets by including the
color-octet contribution. Cheung, Keung, and Yuan calculated the
color-octet $J/\psi$ production in the $\Upsilon$
decay~\cite{Cheung:1996mh}.

According to the NRQCD factorization formula,
only the leading order (LO) contributions are considered for the
processes $\psi(\Upsilon)\to l^+l^-(c\bar{c})gg$.
In the next-to-leading order (NLO), the  decay rate receives
relativistic corrections, whose long-distance
matrix elements are suppressed by $v^2$ compared with the LO
contribution. Notice that the relativistic corrections to the decay
rates in the processes $J/\psi \to \gamma gg$ and $J/\psi \to ggg$
are extremely large and significant~\cite{Keung:1982jb}. One
may expect that the decay rates of the processes $\psi(\Upsilon) \to
l^+l^- gg$ and $\Upsilon\to c\bar{c}gg$ also receive considerable
contributions from the relativistic corrections since those
processes possess similar Feynman diagrams. However, until now, a
thorough analysis including the contributions of the NLO NRQCD
matrix elements is lacking. In this paper, we analyze the decay rate for
$\psi(\Upsilon) \to l^+l^-(c\bar{c})gg$ up to the NLO in the
relativistic expansion in the framework of the NRQCD factorization
formula. We calculate the short-distance coefficients of both the
LO and the NLO NRQCD matrix elements at the tree level
and present the analytic expressions for the distribution of the
invariant mass of the lepton pair or the charm pair. With these
expressions, we are able to study the relativistic corrections in
different kinematic regions, and then provide theoretical discussions.
We also investigate the momentum distribution of the charm quark in our work.
With convolution of a charm quark
fragmenting into a charmed hadron, we are able to predict the momentum
distribution of the charmed hadron.
Since the treatments for inclusive lepton pair production and charm
pair production are quite similar, we concentrate on dealing with
the process of $H(^3S_1)\to l^+l^-gg$. The decay rate of
$\Upsilon\to c\bar{c}gg$ is readily obtained by multiplying a color
factor and substituting the electromagnetic coupling constant
$\alpha$ and the lepton mass $m_l$ into the strong coupling
constant $\alpha_{s}$ and the charm mass $m_c$, respectively.

The remainder of this paper is organized as follows. In
Sec.~\ref{sec:factorization:matrix elements}, we present the NRQCD
factorization formula for the differential decay rate of the process
$H(^3S_1)\to l^+l^-gg$ up to the NLO in $v$.
In Sec.~\ref{sec:k-p}, given the notations and kinematic variables
used in our calculation, we present
the formulas for the differential decay rate as well
as the total decay rate.
Section~\ref{sec:QCD-matching} is devoted to determining the
short-distance coefficients corresponding to the LO and the NLO
NRQCD matrix elements. In Sec.~\ref{sec:conclusion}, we present the
numerical results and provide discussions. A summary is given in
Sec.~\ref{sec:summary}.

\section{NRQCD factorization formula for quarkonium decay process
$H(^3S_1) \to l^+l^- gg $}
\label{sec:factorization:matrix elements}

According to the NRQCD factorization formula, up to relative order
$v^2$, the differential decay rate for a quarkonium $H$ decay
into a lepton pair and light hadrons
can be expressed as~\cite{Bodwin:1994jh}
\begin{equation}
d\Gamma[H({}^3S_1) \to l^+l^- + X] = \frac{dF({}^3S_1)}{  m^2}
\,
        \langle H | {\mathcal O}_1({}^3S_1) |H \rangle
+\frac{dG({}^3S_1)} { m^4}
        \langle H | {\mathcal P}_1({}^3S_1) |H \rangle\;, \label{factor}
\end{equation}
where $m$ signifies the mass of a heavy quark in $H$,
 $\langle H | {\mathcal O}_1({}^3S_1) |H \rangle $ and
 $\langle H | {\mathcal P}_1({}^3S_1) |H \rangle$ are the
 NRQCD matrix elements, and $F({}^3S_1)$ and  $G({}^3S_1)$ are
 the corresponding  short-distance coefficients,
 respectively. Here $H$ can be either charmonium or bottomonium.
The four-fermion operators ${\mathcal O}_1({}^3S_1)$ and ${\mathcal
P}_1({}^3S_1)$ are defined as
\begin{subequations}
\label{3s1-ops}
\begin{eqnarray}
{\mathcal O}_1({}^3S_1)&=&\psi^\dagger\bm{\sigma}\chi\cdot
\chi^\dagger\bm{\sigma}\psi,\\
{\mathcal P}_1({}^3S_1)&=&\frac{1}{
2}\left[\psi^\dagger\bm{\sigma}\chi\cdot \chi^\dagger
\bm{\sigma}(-\frac{i}{2} \tensor{\bm D})^2\psi+
\psi^\dagger\bm{\sigma}(-\frac{i}{2} \tensor{\bm D})^2\chi\cdot
\chi^\dagger \bm{\sigma}\psi \right],
\end{eqnarray}
\end{subequations}
where $\psi$ and $\chi$ are Pauli spinor fields for annihilating a
heavy quark, and creating a heavy antiquark, respectively;
$\sigma^i$ denotes the Pauli matrix; and $\tensor{\bm{D}}$ is
the spatial part of the antisymmetrical covariant derivative:
$\psi^\dagger \tensor{\bm{D}}\chi\equiv \psi^\dagger
{\bm{D}} \chi- ({\bm{D}} \psi)^\dagger \chi$. The subscript
1 on the NRQCD operator indicates that it is a color-singlet
operator. According to the velocity-scaling rules given in
Ref.~\cite{Bodwin:1994jh}, the matrix element of the operator ${\mathcal
O}_1({}^3S_1)$  in  the ${}^3S_1$ state is  of order $v^3$ while
that of the operator ${\mathcal P}_1({}^3S_1)$ is  of order $v^5$. The
latter one is suppressed by $v^2$, which represents the NLO
relativistic corrections to the inclusive $H$ decay.

The vacuum-saturation approximation~\cite{Bodwin:1994jh} can be used to
simplify the decay matrix elements in Eq.~(\ref{3s1-ops}). They
read
\begin{subequations}
\label{VS-me}
\begin{eqnarray}
\label{SVS-me}
\langle H | {\mathcal O}_1({}^3S_1) |H \rangle
&=&|\langle 0|\chi^\dagger\bm{\sigma}\cdot\bm{\epsilon}^*
\psi|H\rangle|^2 \equiv \langle {\mathcal{O}}_1\rangle_{H},\\
\label{PVS-me}
\langle H | {\mathcal P}_1({}^3S_1) |H \rangle&=&
\textrm{Re\,} \big[
\langle H|\psi^\dagger\bm{\sigma}\cdot\bm{\epsilon}\chi|0\rangle
\langle 0|\chi^\dagger\bm{\sigma}\cdot\bm{\epsilon}^*
(-\tfrac{i}{2}\tensor{\bm{D}})^2\psi|H\rangle
\big].
\end{eqnarray}
\end{subequations}
This approximation  is valid up to corrections of relative order $v^4$.
For  convenience, we introduce a dimensionless ratio of the vacuum
matrix elements in Eq.~(\ref{VS-me})
for later use~\cite{Gremm:1997dq,Bodwin:2006dn}:
\begin{eqnarray}
\label{me-ratios}
\langle v^2\rangle_{H}&=&
\frac{\langle 0|\chi^\dagger\bm{\sigma}\cdot\bm{\epsilon}^*
(-\tfrac{i}{2}\tensor{\bm{D}})^{2}\psi|H\rangle}
{{m}^2\langle 0|
\chi^\dagger\bm{\sigma}\cdot\bm{\epsilon}^*\psi
|H\rangle}.
\end{eqnarray}
This quantity characterizes the typical size of relativistic
corrections for $H$.

Equation~(\ref{factor}) implies that, to predict the decay rate, one
needs to determine both the short-distance coefficients and the
NRQCD matrix elements.  The NRQCD matrix elements have been
extensively studied  by means of  lattice QCD~\cite{Bodwin:1993wf},
the nonrelativistic quark model~\cite{Eichten:1995}, and fitting the
experimental data~\cite{Bodwin:2007fz, Guo:2011tz}. Therefore, once we
determine the short-distance coefficients $F({}^3S_1)$ and
$G({}^3S_1)$, with those values of matrix elements, we may calculate
the differential decay rate in Eq.~(\ref{factor}).  To determine
the $F({}^3S_1)$ and $G({}^3S_1)$ at the tree level,  we apply the
factorization formula to the process of an on-shell $Q\bar{Q}$ pair
near the threshold in a spin-triplet and color-singlet state
decaying to $l^+l^-gg$:
\begin{eqnarray}
d\Gamma[Q\bar{Q}_1({}^3S_1) \to l^+l^- gg] &=& \frac{dF({}^3S_1)}{
 m^2} \,
        \langle Q\bar{Q}_1({}^3S_1) | {\mathcal O}_1({}^3S_1) |
        Q\bar{Q}_1({}^3S_1) \rangle  \nonumber\\
&&\hbox{}+ \frac{dG({}^3S_1)}{  m^4}
        \langle Q\bar{Q}_1({}^3S_1)
        | {\mathcal P}_1({}^3S_1) |Q\bar{Q}_1({}^3S_1) \rangle\;.
\label{factor-ccbar}
\end{eqnarray}
Notice that the factorization formula (\ref{factor-ccbar}) takes a
similar form to (\ref{factor}) except that the hadron state is
substituted into the on-shell free quark pair state with
the same quantum number as the hadron. The decay rate
in (\ref{factor-ccbar}) can be calculated both in the QCD perturbation theory
and in the NRQCD factorization formula. By matching both sides, the
short-distance coefficients can then be determined.

\section{Kinematics and Formulas for the decay rate}
\label{sec:k-p}

\subsection{Kinematics and definitions}
\label{sec:kinematics}

In this section, we define notations for the kinematics involved in our work.
We take $p_1$ and $p_2$ to be the momenta of the incoming
heavy quark $Q$ and heavy antiquark $\bar{Q}$, respectively, which
are on their mass shells: $p_1^2=p_2^2=m^2$. They are expressed as
linear combinations of the total momentum $P$ and half of their
relative momentum $q$:
\begin{subequations}
\label{momenta}
\begin{eqnarray}
p_1&=&P/2+q,\\
p_2&=&P/2-q.
\end{eqnarray}
\end{subequations}

In the center of mass frame of the quarkonium, the momenta are given by
\begin{subequations}
\label{rest-frame}
\begin{eqnarray}
P&=&(2E,\bm{0}),\\
q&=&(0,{\bm q}),
\end{eqnarray}
\end{subequations}
where the orthogonal relation $P\cdot q=0$ is satisfied.

We also assign $k_1, k_2$ to be the momenta of the two final-state gluons, and
$l_1, l_2$ to be the momenta of the produced lepton pair. Therefore, the momentum $Q$
of the virtual photon yields to $Q=l_1+l_2$.
These momenta satisfy
\begin{subequations}
\label{rest-frame-k}
\begin{eqnarray}
k_1^2= k_2^2&=&0,\\
l_1^2 = l_2^2&=&m_l^2,
\end{eqnarray}
\end{subequations}
where $m_l$ denotes the mass of the lepton.

For convenience, we introduce a set of dimensionless variables
\begin{subequations}
\begin{eqnarray}
x_1&=&\frac{2k_1\cdot P}{P^2}, \,\, x_2=\frac{2k_2\cdot P}{P^2},
\,\, x_3=\frac{2Q\cdot P}{P^2}, \,\, z=\frac{Q^2}{P^2},\\
r&=&\frac{4m_l^2}{P^2}, \,\,\,\,
y_1=\frac{|{\bm l}_1|}{|{\bm l_1}|_{max}}=\frac{|{\bm l}_1|}{m_l}\sqrt{\frac{r}{1-r}},
\label{definition}
\end{eqnarray}
\end{subequations}
where $y_1$ represents the momentum fraction for the lepton and $|{\bm l_1}|_{max}$ denotes
the maximum of the lepton momentum in the quarkonium center of mass frame.
In the following subsection, we will show that all
the involved Lorentz invariant kinematic quantities can
be rewritten in terms of these new variables.

\subsection{The formulas for the decay rate}
\label{phase space:fb}
\subsubsection{Differential decay rate of the invariant mass of the lepton pair}
\label{sec:dif-z}
For the decay process $H({}^3S_1)(P)\to l^+(l_1)l^-(l_2)g(k_1)g(k_2)$,
it involves a four-body phase space integral, which can be expressed as
\begin{eqnarray}
\int \!\! d\phi_4&=& \int\!\! \frac{d^3k_1}{ (2\pi)^{3}2k^{0}_1}
\frac{d^3k_2}{ (2\pi)^{3}2k^{0}_2} \frac{d^3l_1}{
(2\pi)^{3}2l^{0}_1} \frac{d^3l_2}{ (2\pi)^{3}2l^{0}_2}
(2\pi)^4\delta^4(P-k_1-k_2-l_1-l_2). \label{eq:phase-four-1}
\end{eqnarray}
Since there is no divergence emerging in our calculation, dimensions
of the space-time are set to $4$ in
(\ref{eq:phase-four-1}). In order to compute the invariant mass
distribution of the lepton pair, we decompose  the four-body phase space integral
(\ref{eq:phase-four-1}) into the product of a
two-body phase space integral for the lepton pair and a three-body one
by inserting the following two identities:
\begin{eqnarray}
 \int\!\! \frac{d^4Q}{ (2\pi)^4}(2\pi)^4\delta^4(Q-l_1-l_2)=1,
\ \ \ \ \ P^2\int\!\! \frac{dz}{ 2\pi}2\pi\delta(Q^2-P^2 z)=1.
\label{eq:two-id}
\end{eqnarray}
After integrating out the energy $Q^0$ through the delta function, we
get
\begin{eqnarray}
 \int \!\! d\phi_4&=&  \int\!\!\frac{dz}{ 2\pi}
 \int \!\! d\phi_3 \int \!\! d\phi_2,
\label{eq:phase-four-2}
\end{eqnarray}
where $\int \!\! d\phi_2$ and $\int \!\! d\phi_3$ are expressed  as
\begin{subequations}
\begin{eqnarray}
\int \!\! d\phi_2&=& P^2 \int\!\! \frac{d^3l_1}{ (2\pi)^{3}2l^{0}_1}
\frac{d^3l_2}{
(2\pi)^{3}2l^{0}_2}(2\pi)^4\delta^4(Q-l_1-l_2),\\
\label{two-body} \int \!\! d\phi_3&=&\int\!\!\frac{d^3Q}{
(2\pi)^{3}2Q^{0}} \frac{d^3k_1}{ (2\pi)^{3}2k^{0}_1} \frac{d^3k_2}{
(2\pi)^{3}2k^{0}_2}(2\pi)^4\delta^4(P-k_1-k_2-Q). \label{three-body}
\end{eqnarray}
\end{subequations}

On the other side, the squared amplitude can be expressed
as the contraction of the
leptonic tensor ${\widetilde L}^{\mu\nu}$ and hadronic tensor
${\widetilde H}_{\mu\nu} $:
\begin{eqnarray}
|{\mathcal M}|^2={\widetilde L}^{\mu\nu}{\widetilde H}_{\mu\nu},
\label{eq:amps}
\end{eqnarray}
where the leptonic tensor ${\widetilde L}^{\mu\nu}$ is given by
\begin{eqnarray}
{\widetilde L}^{\mu\nu}=\frac{e^2}{Q^4}{\rm
Tr}[(\not\!l_1+m_l)\gamma^\mu(\not\!l_2-m_l)\gamma^\nu].
\label{eq:lepton1}
\end{eqnarray}
It follows after integrating over the phase space momenta that
\begin{eqnarray}
L^{\mu\nu}\equiv\int\!\!  d\phi_2{\widetilde L}^{\mu\nu}
=\bigg(\!\!-g^{\mu\nu}+\frac{Q^\mu Q^\nu}{Q^2}\bigg)\times L,
\label{eq:lepton2}
\end{eqnarray}
where the Lorentz invariant $L$ is given by
\begin{equation}
L= \frac{2\alpha}{ 3 z} \ \sqrt{1-\frac{r}{ z}}\ (1+\frac{r}{ {2
z}}), \label{eq:lepton3}
\end{equation}
where $\alpha$ is the fine structure constant.
As a result, we are able to write the decay rate as
\begin{equation}
\Gamma=\frac{1}{2}\int\!\! \frac{dz}{2\pi}L\int\!\! d\phi_3 {\widetilde H}_{\mu\nu}
\bigg(-g^{\mu\nu}+\frac{Q^\mu Q^\nu}{Q^2}\bigg),
\label{eq:width1}
\end{equation}
where the factor $\frac{1}{2}$ accounts for the indistinguishability of
the two gluons in the final states. It is not hard to
find that the second
term in the parentheses of (\ref{eq:width1}) does not contribute
due to the current conservation.

The three-body phase space integral $\int\!\! d\phi_3$ can
generically be
expressed as the integral of two dimensionless variables $x_1$ and
$x_2$:
\begin{eqnarray}
\label{eq:phase-three-1} \int \!\! d\phi_3&=& \frac{P^2}{
128\pi^3}\int\!\! dx_1 dx_2.
\end{eqnarray}
Up to now, we have reduced
the four-body phase space integral~(\ref{eq:phase-four-1}) into the integration over
three variables: $z$, $x_1$, and $x_2$. The corresponding boundaries for these
variables are given by
\begin{eqnarray}
\frac{1-x_1-z }{ 1-x_1}\ge  x_2 \ge 1-x_1-z ,\ \ \ 1-z\ge x_1 \ge 0, \ \ \  1\ge z \ge r.
 \label{integral:limit}
\end{eqnarray}
To simplify further the calculation, we make the variable transformation:
\begin{subequations}
\begin{eqnarray}
x_1&=&(1-z)x,\\
x_2&=&\frac{(1-z)(1-x)[1-(1-z)x y]}{1-(1-z)x}.
\end{eqnarray}
\end{subequations}
After this transformation,
the area of the integration is significantly simplified as
\begin{eqnarray}
&&1\ge x \ge 0, \,\,\,\,\,\,\,\,\,\,\,\,\, 1 \ge y \ge 0.
\end{eqnarray}

Now, the expression of the decay rate reduces to
\begin{equation}
\Gamma=\frac{1}{2}\frac{P^2}{(4\pi)^4}
\int^1_r  dz \int^1_0 dx  \int^1_0 dy \frac{(1-z)^3(1-x)x}{1-(1-z)x}
L\times(-g^{\mu\nu}){\widetilde H}_{\mu\nu}.
\label{eq:width-2}
\end{equation}
From
(\ref{eq:width-2}), we notice the
principal task is to analyze
the subprocess $H(^3S_1)\to \gamma^*gg$ (corresponding to the contribution
from the hadron part
${\widetilde H}_{\mu\nu}g^{\mu\nu}$). In the next section, we will use Eq.~(\ref{eq:width-2})
to evaluate the total decay rate as well as the differential decay rate
over the invariant mass of the lepton pair, equivalently, the
dimensionless variable $z$.

\subsubsection{Momentum distributions of the charm quark and
the charmed hadron}
\label{sec:dif-L}
In this section, we first derive the formulas to calculate the momentum distribution
of the charm quark in the decay process $\Upsilon \to g^*gg \to c\bar{c} gg$.
The momentum distribution of a charmed hadron $h$ is then obtained
by convolving it with a
fragmentation function, which describes a charm quark fragmentation into the
meson $h$.

As introduced in Sec.~\ref{sec:dif-z},
we decompose the phase space integration into two parts by
inserting the identities (\ref{eq:two-id}).
Since we want to observe the momentum distribution
of the charm quark, we can integrate out the momenta of the two final-state gluons. To this end, we introduce a tensor $T^{\mu\nu}$ which depends
only on the momenta $P$ and $Q$ as
\begin{eqnarray}
T^{\mu\nu}&\equiv&\int\!\! {d^3k_1\over (2\pi)^{3}2k^{0}_1} {d^3k_2\over
(2\pi)^{3}2k^{0}_2}(2\pi)^4 \delta^4 (P-Q-k_1-k_2){\widetilde H}^{\mu\nu}\nonumber\\
 &=& (-g^{\mu\nu}+
\frac{Q^\mu Q^\nu}{Q^2})H_1+\frac{1}{P^2}(P^\mu-
Q^\mu \frac{ P\cdot Q}{Q^2})(P^\nu-
Q^\nu \frac{ P\cdot Q}{Q^2})H_2,
\label{form-factor-1}
\end{eqnarray}
where $H_1$, $H_2$
are Lorentz invariant form factors. In the last step of (\ref{form-factor-1}),
we have applied the Lorentz covariance and current conservation.
By contracting $g^{\mu\nu}$ and $P^{\mu}P^{\nu}$ separately
in (\ref{form-factor-1}), we are able to obtain the expressions of these two form factors.
We notice that
$H_1$ and $H_2$ are independent on the momenta of the two final fermions.
To obtain the decay rate, we need to include the  charm quark pair part
as well as the remaining phase space.

Contracting with the leptonic tensor,\footnote{Here we should replace
the leptonic tensor ${\widetilde L}^{\mu\nu}$ in (\ref{eq:lepton1}) with 
the corresponding tensor for the charm quark pair; however, we still use 
(\ref{eq:lepton1}) to implement the calculation and the difference
will be compensated by multiplying a factor.}
we readily obtain
\begin{eqnarray}
{\widetilde T}&\equiv& T^{\mu\nu}{\widetilde L}_{\mu\nu}=
\int\!\! {d^3k_1\over (2\pi)^{3}2k^{0}_1} {d^3k_2\over
(2\pi)^{3}2k^{0}_2}(2\pi)^4 \delta^4 (P-Q-k_1-k_2)\times
\frac{2\pi r}{m_l^2}\times T\nonumber\\
&=&\frac{2\pi r\alpha}{m_l^2z^{2}}\bigg\{
   (r+2 z)H_1-H_2 \bigg[(1-r)
   y_1^2+r+z-x_3 \sqrt{(1-r)
   y_1^2+r}\bigg]\bigg\}.
\label{T}
\end{eqnarray}

Now, we turn to carry out the two phase space
integration $\int\!\! d\phi_2$ and $\int\!\! d\phi_3$. For $\int\!\! d\phi_2$,
we have
\begin{eqnarray}
\int\!\!\! d\phi_2&=&P^2\int\!\! {d^3l_1\over (2\pi)^{3}2l^{0}_1} {d^3l_2\over
(2\pi)^{3}2l^{0}_2}(2\pi)^4\delta^4(l_1+l_2-Q)\nonumber\\
&=&\frac{P^2}{8\pi}\int\!\!\frac{|{\bm l}_1|d|{\bm l}_1|}
{\sqrt{{\bm l}_1^2+m_c^2}|\bm Q|}\nonumber\\
&=&\frac{m_l^2}{2\pi r}\int\!\! \frac{y_1dy_1}{\sqrt{(1-r)y_1^2+r}\sqrt{x_3^2-4z}},
\label{form-factor-2}
\end{eqnarray}
with the boundaries of $y_1$:
\begin{eqnarray}
y_{1+}
\ge y_1\ge |y_{1-}|,
\label{y1-bound-1}
\end{eqnarray}
where
\begin{eqnarray}
y_{1\pm}=\frac{x_3}{2\sqrt{1-r}}\bigg(\sqrt{1-\frac{4z}{x_3^2}}\pm
\sqrt{1-\frac{r}{z}}\bigg).
\label{y1-bound-2}
\end{eqnarray}

We then deal with the phase space integral $\int\!\! d\phi_3$.
Analogously, we can reduce the integral $\int\!\! d\phi_3$ into
(\ref{eq:phase-three-1}).
Nevertheless,
since the boundaries (\ref{y1-bound-2}) contain $x_3$, we prefer to
choose another set of integration variables, such as $x_1$ and $x_3$:
\begin{eqnarray}
\label{eq:phase-three-1-1}
\int \!\! d\phi_3&=& {P^2 \over 128\pi^3}\int \!\! dx_3 dx_1.
\end{eqnarray}
The corresponding boundaries of $x_3$ and $x_1$ are
\begin{subequations}
\begin{eqnarray}
&&1+z \ge x_3\ge 2\sqrt{z},\\
&&x_{1+}\ge x_1\ge x_{1-},
\label{bound-x3-0}
\end{eqnarray}
\end{subequations}
where
\begin{eqnarray}
x_{1\pm}=\frac{1}{2}\bigg(2-x_3\pm\sqrt{x_3^2-4z}\bigg).
\label{bound-x1}
\end{eqnarray}

In addition, as shown in (\ref{eq:phase-four-2}),
to get the decay rate,
we should include another integration
over $z$. The corresponding boundaries of $z$ are shown in
(\ref{integral:limit}) to be $1\ge z\ge r$.

Finally, the decay rate can be expressed as
\begin{eqnarray}
\Gamma=\frac{1}{2}\frac{P^2}{(4\pi)^4}\int\!\! dzdx_3dx_1dy_1 \frac{y_1}{\sqrt{(1-r)y_1^2+r}\sqrt{x_3^2-4z}}\times T,
\label{dif-integration}
\end{eqnarray}
where $T$ is defined in (\ref{T}).

In order to get the momentum distribution,
we need to change the integration order in (\ref{dif-integration}), and to make $y_1$
be the last integral. Notice that the boundaries of $y_1$ are independent
of
$x_1$; we need not change the order of the integration of $x_1$.
This calculation is tedious but straightforward.
Here we present the expression as follows:
\begin{eqnarray}
\Gamma&=&\frac{1}{2}\frac{P^2}{2(4\pi)^5}\bigg(\int_0^{\frac{\sqrt{1-r}}{2}}\!\! dy_1\int_r^{z_-}\!\!dz\int_{x^\prime_{3-}}^{x^\prime_{3+}}\!\!dx_3+\int_0^1\!\! dy_1\int_{z_-}^{z_+}\!\!dz\int_{x^\prime_{3-}}^{1+z}\!\!dx_3\bigg)
\int_{x_{1-}}^{x_{1+}}dx_1\nonumber\\
&&\times \frac{y_1}{\sqrt{(1-r)y_1^2+r}\sqrt{x_3^2-4z}}\times T,
\label{dif-integration-1}
\end{eqnarray}
where the boundaries of $x_1$ are given in (\ref{bound-x1}), and
\begin{eqnarray}
&&x^\prime_{3\pm}=\frac{2}{r}\bigg(z\sqrt{(1-r)y_1^2+r}\pm y_1\sqrt{(1-r)(z-r)z}\bigg).
\label{bound-x3}
\end{eqnarray}
In addition, the boundaries of variable $z$ are the positive solution of
the following equation:
\begin{eqnarray}
\frac{(1+z_{\pm})\sqrt{1-\frac{r}{z_{\pm}}}\mp(1-z_{\pm})}{2\sqrt{1-r}}=y_1.
\label{bound-z}
\end{eqnarray}
With the formula (\ref{dif-integration-1}), and the boundaries
(\ref{bound-x1}) (\ref{bound-x3}) (\ref{bound-z}), we can
carry out a calculation of the distribution of the charm quark momentum fraction $y_1$.
Now,
we go further to investigate the charmed-hadron momentum distribution.
As discussed in Ref.~\cite{Bodwin:2007fz}, the momentum
distribution of a charmed hadron produced in $\Upsilon$ decay is softer
than that of the charm, due to the effect of
hadronization. The momentum distribution of a charmed hadron $h$ can be
obtained by convolving the charm momentum distribution with a fragmentation
function for the charm quark fragmentation into the $h$.

The fragmentation function $D_{c\to h}(z^\prime)$ describes the probability
of a charm quark with light-cone momentum
$l_1^0+|{\bm l_1}|$ hadronizing into a charmed hadron $h$ with
light-cone momentum $l^0_h+|{\bm l_h}|=z^\prime(l_1^0+|{\bm l_1}|)$.
The fraction $z^\prime$ can be
expressed in terms of scaled light-cone momentum fractions
$z_1$ for the charm and $z_h$ for the charmed hadron, which are
analogous to the scaled momenta $y_1$ and
$y_h$~\cite{Bodwin:2007zf}, where $z_1$ is
\begin{eqnarray}
z_1 = \frac{\sqrt{(1-r)y_1^2 + r} + \sqrt{1-r} \, y_1}
           {1 + \sqrt{1-r}}.
\label{z1}
\end{eqnarray}
Then, the fraction $z^\prime$ is expressed as
\begin{eqnarray}
z^\prime = \frac{z_h}{z_1}\times
\frac{\left. (l^0_h+|{\bm l_{h}}|)\right|_\textrm{max}}
     {\left. (l^0_1+|{\bm l_{1}}|)\right|_\textrm{max}},
\label{zprime}%
\end{eqnarray}
where the last factor on the right-hand side of Eq.~(\ref{zprime}) becomes
unity if the difference between the mass of the charm quark
and that of the charmed hadron can be neglected.
With this approximation, the momentum distribution of
the charmed hadron can be written as~\cite{Bodwin:2007fz}
\begin{eqnarray}
\frac{d\Gamma}{d y_h}
&=&
\frac{dz_h}{dy_h} \int_{z_h}^{z_m} \frac{dz_1}{z_1} \,
D_{c\to h}(z_h/z_1)\, \frac{dy_1}{dz_1}
\frac{d\Gamma}{d y_1}
\nonumber\\
&=&
\frac{\sqrt{1-r}}{\sqrt{(1-r)y_h^2+r }}
\int_{y_h}^{y_m} dy_1
  \mathcal{D}_{c\to h}\left(
  \frac{\sqrt{(1-r)y_h^2+r} +\sqrt{1-r}y_h }
       {\sqrt{(1-r)y_1^2+r} +\sqrt{1-r}y_1 }
\right) \frac{d\Gamma}{dy_1},
\label{dGamD}%
\end{eqnarray}
where $\mathcal{D}_{c\to h}(z^\prime)= z^\prime D_{c\to h}(z^\prime)$, $y_m$
represents the upper boundary for $y_1$ in (\ref{dif-integration-1}),
which equals $\sqrt{1-r}/2$ and 1 corresponding to
the first term and the second term in the parentheses, and $z_m$ corresponds
to the value of $z_1$ when $y_1$ takes $y_m$ in (\ref{z1}).

The formulas (\ref{dif-integration-1}) (\ref{dGamD}),
and the boundaries
(\ref{bound-x1}) (\ref{bound-x3}) (\ref{bound-z}) can be used to
carry out a calculation of the distribution of the charmed-hadron momentum
fraction $y_h$.
In Sec.~\ref{sec:conclusion}, we will utilize these formulas to
make predictions.

\section{Matching the short-distance coefficients up to NLO in $v$}
\label{sec:QCD-matching}
In this section, we determine the differential short-distance coefficients
$dF(^3S_1)$ and $dG(^3S_1)$ that appeared in (\ref{factor}). The short-distance coefficients
are then readily obtained by integrating over the integration variables.
Now, we describe the strategy.
By employing the formulas derived in the previous section,
we first
calculate the differential decay rate for the process of a
color-singlet spin-triplet $S$-wave heavy quark pair decay into a lepton pair plus
two gluons
$Q{\bar
Q}_1(^3S_1)\to l^+l^-gg$ in the QCD perturbation theory, up to
the NLO in $v$, and then carry out the differential decay rate of
the same process
in the NRQCD factorization
formula. Finally, the short-distance coefficients $dF(^3S_1)$
and $dG(^3S_1)$ in (\ref{factor}) are immediately determined
by identifying these two calculations.

\subsection{Amplitude of $Q\bar{Q}\to \gamma^*gg$}
\label{amplitude of QQ}

\begin{figure}[ht]
\begin{center}
\includegraphics*[scale=0.8]{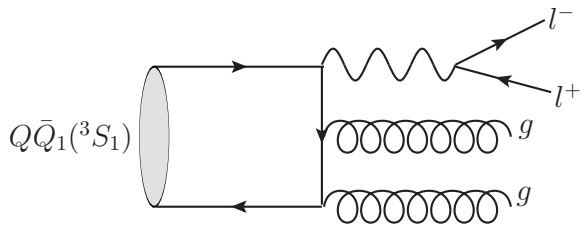}
\caption{The tree-level Feynman diagrams for $Q\bar{Q}_1 ({}^3S_1)
\to l^+l^-  g g$. For simplicity, the crossed diagrams have been
suppressed.} \label{feynman:fig}
\end{center}
\end{figure}

As we have demonstrated in (\ref{eq:lepton3}) (\ref{eq:width-2})
(\ref{T}), and (\ref{dif-integration-1}),
the lepton part has been explicitly
written out. We still have to deal with
the subprocess $Q(p_1)\bar{Q}(p_2)\to
\gamma^*(Q)g(k_1)g(k_2)$. At the tree level, there are 6 diagrams
contributing to the amplitude as shown in Fig.~\ref{feynman:fig}.
Given the momenta
defined in Sec.~\ref{sec:kinematics}, the amplitude of the process
reads
\begin{equation}
\label{A-s1s2}
 A(s_1,s_2)\;=\;\bar{v}(p_2,s_2)\; T_{\mu}\; u(p_1,s_1),
\end{equation}
where $u(p_1,s_1)$ and $ v(p_2,s_2) $ are the spinors of the heavy quark and
antiquark, respectively, and $T_{\mu}$ represents the products of  Dirac
matrices and color-space matrices. According to Fig.~\ref{feynman:fig},
the
expression of $T_{\mu}$ reads
\begin{eqnarray}
T_{\mu}&=& (- i e_Q e g_s^2)\ T^b T^a \otimes
\not\!\epsilon_2^*(k_2) \frac{1}{{\not\!k_2 - \not\! p_2-m}}\not\!
\epsilon_1^{*}(k_1)\frac{1}{{\not\!k_1+\not\!k_2-\not\!
p_2-m}}\gamma_\mu
  + 5 \ \textrm{perms},
\label{amp:ex}
\end{eqnarray}
where $e, g_s$ denote the QED and QCD coupling constant,
respectively, $e_Q$ denotes the electric charge number of the heavy
quark, $a, \epsilon_1$ and $b, \epsilon_2$ represent the color
indices and the polarization vectors of the two gluons, and $\mu$
corresponds to the Lorentz index of the virtual photon.


\subsection{Projection of spin-triplet $Q\bar{Q}$ state}
\label{projection:spin-singlet}

The amplitude given in (\ref{A-s1s2}) describes the decay of the
heavy quark and the antiquark state with the spins of the third
component $s_1$ and $s_2$, respectively. To calculate the decay of
the $Q\bar{Q}$ in the spin-triplet state and color-singlet state,
one needs to project the total spin state of the $Q\bar{Q}$  pair onto
the spin-triplet and color-singlet $Q(p_1)\bar{Q}(p_2)$ state. This
can be done by introducing the projection operator
$\Pi_3(p_1,p_2)$\cite{Bodwin:2002hg} expressed by
\begin{eqnarray}
\Pi_3(p_1,p_2)&=& \sum_{s_1,s_2} u(p_1,s_1) \bar{v}(p_2,s_2)
\langle \frac{1}{ 2},s_1;\frac{1}{ 2},s_2|1 \epsilon \rangle \otimes
\frac{\mathbf{1}_c}{ \sqrt{N_c}}
\nonumber\\
&=&- \frac{1 }{ 8\sqrt{2}E^2 (E + m)}
(/\!\!\!{p_1}+m)(\,/\!\!\!P\!+\!2E ) \,/\!\!\!\epsilon
(/\!\!\!p_2-m) \otimes  \frac{\mathbf{1}_c}{ \sqrt{N_c}},
\label{spin:singlet:projector}%
\end{eqnarray}
where $\mathbf{1}_c$ is the unit matrix in the fundamental
representation of the color SU(3) group, and $\epsilon$ is the
polarization vector of the spin-triplet state. The above spin-triplet
projector is derived by assuming the nonrelativistic normalization
convention for Dirac spinor. With this projection operator,  the
amplitude for a spin-triplet and color-singlet $Q\bar{Q}$ pair
annihilation decay reads
\begin{eqnarray}
 {\mathcal A}_{\mu}^{\rm sing}[Q\bar{Q}\to \gamma^*gg] =
\textrm{Tr}\bigg\{ \Pi_3(p_1,p_2) T_{\mu}
\bigg\},
\label{spin:triplet:ampltude}%
\end{eqnarray}
where the trace is understood to act on both Dirac and color spaces.

\subsection{Projection of $S$-wave amplitude}
\label{sec:proj}

Besides projecting the $Q\bar{Q}$ pair onto  the spin-triplet state,
to account for the contribution from the $S$-wave
orbital-angular-momentum state, one has to project further the
$Q\bar{Q}$ state onto the $S$-wave state. It can be done by
averaging the amplitude ${\mathcal A}^{\rm sing}$ over all
directions of the relative momentum $\mathbf{q}$ in the $Q\bar{Q}$
rest frame.

The amplitude can be  expanded  in  terms of the powers of ${\bm
q}^2$ and the series can be truncated to the desired order. Since here
we are only interested in the NLO relativistic corrections,
we may do it by  expanding the spin-triplet amplitude ${\mathcal
A}^{\rm sing}$ in $q^\mu$ through quadratic order, then making the
following replacement~\cite{Bodwin-Lee:2003}:
\begin{eqnarray}
q^\mu q^\nu  &\to&  \frac{\mathbf{q}^2 }{ 3}\: \Pi^{\mu\nu}(P),
\label{S-wave:projection}%
\end{eqnarray}
where
\begin{eqnarray}
\Pi^{\mu\nu}(P) &\equiv& -g^{\mu\nu}+\frac{P^\mu P^\nu}{ P^2}.
\end{eqnarray}
\subsection{The decay rate of $Q\bar{Q}_1({^3S_1})\to l^+l^-gg$
up to relative order $v^2$}
Since the calculation for the momentum
distributions of the charm quark and charmed hadron are similar to
that of the invariant
mass distribution of the lepton pair, in the following subsections,
we merely demonstrate
the latter.

We now proceed to compute the lepton pair invariant mass distribution for the process
$Q \bar{Q}_1 ({}^3S_1)\to l^+l^-
gg$ at the LO and the NLO in $v$, based on the techniques described in
Sec.~\ref{sec:proj}.

We
first expand the amplitude given in (\ref{spin:triplet:ampltude}) in
terms of  ${\bm q}$ up to quadratic order, then apply
(\ref{S-wave:projection}) to extract the $S$-wave part
\begin{eqnarray}
{\overline A}_\mu={\overline A}^{(0)}_\mu+{\overline A}_{\mu}^{(2)}
\frac{{\bm q}^2}{m^2}+
{\mathcal O}({\bm q}^4).
\end{eqnarray}

The hadronic tensor in (\ref{eq:width-2}) is then given by squaring
the amplitude ${\overline A}_\mu$, averaging over the polarizations
of the initial state, and  summing over the polarizations of the two
gluons:
\begin{eqnarray}
{\widetilde H}_{\mu\nu}=\frac{1}{3}\sum_{\rm{pol}}{\overline A}_\mu{\overline A}^*_\nu.
\end{eqnarray}
Substituting it into (\ref{eq:width-2}), the decay rate is expressed as
\begin{eqnarray}
\Gamma&=&-\frac{4E^2}{2(4\pi)^4}\int\!\! dzdxdy \frac{(1-z)^3(1-x)x}{1-(1-z)x}\times
L\times \frac{1}{3}\sum_{pol}{\overline A}^\mu
{\overline A}^*_\mu \nonumber\\
&=&-\frac{2m^2}{3}\frac{1}{(4\pi)^4}\int\!\! dzdxdy
\frac{(1-z)^3(1-x)x}{1-(1-z)x}\times
L\nonumber\\
&&\times \sum_{pol}\bigg[{\overline A}^{(0)\mu}
{\overline A}^{(0)*}_\mu
+\bigg({\overline A}^{(0)\mu}
{\overline A}^{(0)*}_\mu+2{\textrm Re}[{\overline A}^{(0)\mu}
{\overline A}^{(2)*}_\mu]\bigg)
\times\frac{{\bm q}^2}{m^2}+{\mathcal O}({\bm q}^4)\bigg]\;.
\label{eq:width-3}
\end{eqnarray}

In the calculation, we employ the mathematica package 
Feyncalc~\cite{Mertig:1990an} to implement
the arithmetic of Dirac trace and Lorentz contraction. The resultant
distribution of the invariant mass of the lepton pair reads
\begin{eqnarray}
\label{Gamma-diff-qcd}
\frac{d\Gamma}{ dz}&=& \frac{4 \alpha^2 \alpha_s^2 e_Q^2} { 27\pi
m^2 } \sqrt{1-\frac{r}{ z}}\ (1+ \frac{r }{ 2 z})\;
\bigg(f_0(z)+f_2(z)\frac{{\bm q}^2}{m^2}\bigg),
\end{eqnarray}
where the analytic expressions for
$f_0(z)$ and $f_2(z)$ are given by
\begin{eqnarray}
f_0(z) &=& \frac{4}{z(1-z)^2} \bigg \{ \bigg[(2 z^3-z^2-12 z+8) \tan
   ^{-1}\bigg(\sqrt{\frac{1-z}{z}}\bigg)+2 \sqrt{z(1-z)}
   (4 z^2-9 z+8)\bigg] \nonumber\\
   &\times& \tan
   ^{-1}\bigg(\sqrt{\frac{1-z}{z}}\bigg)-9 (1-z)(z^2-2 z+2)+z (5 z^2-14 z+3) \log z\bigg \},
\label{eq:decayrate-result0}
\end{eqnarray}
and
\begin{eqnarray}
f_2(z)&=& \frac{4}{9z(1-z)^3}\bigg \{ \bigg[3(4 z^4-8 z^3-57 z^2+96
z-38)
\tan^{-1}\left(\sqrt{\frac{1-z}{z}}\right)-6 \sqrt{z(1-z)}\nonumber\\
&\times&(17 z^2-51 z+31)\bigg ]\tan
   ^{-1}\left(\sqrt{\frac{1-z}{z}}\right)-(1-z) (61 z^3-192
   z^2+386 z-198)\nonumber\\
&+&2 z (z^3-55 z^2+43 z-13) \log z\bigg \}.
\label{eq:decayrate-result2}
\end{eqnarray}

Notice that, when extracting the relativistic corrections,
we do not expand $r$ in terms of $E=\sqrt{m^2+{\bm q}^2}$ in
(\ref{Gamma-diff-qcd}).
Actually, from the
expression of (\ref{Gamma-diff-qcd}), we find the
differential decay rate is sensitive to the value of $r$ in the
region of $z\to r$.
Moreover,
the decay rate develops a strong dependence on $r$ from this region,
i.e., 
$\int_r dz \frac{d\Gamma}{dz}\propto \log r$.
In our numerical calculation,
we will choose
$r=4m_l^2/P^2=4m_l^2/m_H^2$,
where $m_H$ is the mass of the initial quarkonium.
Since
the quarkonium mass is well measured, this choice may also
reduce the uncertainties from the input parameters.

For the same reasons, we will make the choice of
$r=4m_D^2/m_H^2$
in (\ref{dif-integration-1})
when evaluating the momentum distributions for the
charm quark and the charmed hadron.

\subsection{The short-distance coefficients $d\bm{F({}^3S_1)}$ and $d\bm{G({}^3S_1)}$}
\label{cal:coeff:F:G}

To determine the short-distance coefficients, we need to calculate
the parton level process $Q\bar{Q}_1 ({}^3S_1) \to l^+l^- g g$ in the NRQCD factorization
formula. The involved  matrix elements are easily obtained by
perturbative NRQCD:
\begin{subequations}
\label{eq:ps-norm}
\begin{eqnarray}
\label{eq:ps0-norm}
\langle Q\bar{Q}_1({}^3S_1)|{\mathcal O}_1(^3S_1)|Q\bar{Q}_1({}^3S_1)
\rangle &=& 2N_c, \\
\label{eq:ps2-norm}
\langle Q\bar{Q}_1({}^3S_1)|{\mathcal P}_1(^3S_1)|Q\bar{Q}_1({}^3S_1)
\rangle&=& 2N_c\, \bm{q}^{2},
\end{eqnarray}
\end{subequations}
where the state of the heavy quark pair is normalized
nonrelativistically, and the factor $2N_c$ accounts for the spin and
color normalization.

Substituting (\ref{eq:ps-norm}) into (\ref{factor-ccbar}), we can
write down the corresponding differential decay rate in the NRQCD factorization
formula:
\begin{eqnarray}
\frac{d}{dz}\Gamma(Q\bar{Q}_1({}^3S_1)\to
l^+l^-gg)=\frac{2N_c}{m^2}\bigg(\frac{d\,F({}^3S_1)}{dz}
+\frac{d\,G({}^3S_1)}{dz}\frac{{\bm q}^2}{m^2}+{\mathcal O}({\bm q}^4)
\bigg). \label{NRQCD width}
\end{eqnarray}
Matching the QCD side and the NRQCD side by equating
(\ref{Gamma-diff-qcd}) with (\ref{NRQCD width}), one determines the
short-distance coefficients $\frac{d\,F({}^3S_1)}{dz}$ and
$\frac{d\,G({}^3S_1)}{dz}$:
\begin{subequations}
\begin{eqnarray}
\frac{d\,F({}^3S_1)}{dz}&=& \frac{2 \alpha^2 \alpha_s^2 e_Q^2
}{ 81\pi  } f_0(z) \sqrt{1-\frac{r}{ z}}\ (1+ \frac{r }{ 2 z}),\\
\frac{d\,G({}^3S_1)}{dz}&=& \frac{2 \alpha^2 \alpha_s^2 e_Q^2
}{81\pi} f_2(z) \sqrt{1-\frac{r}{ z}}\ (1+ \frac{r }{ 2 z}).
\end{eqnarray}
\label{short-diff}
\end{subequations}

\begin{figure}[tb]
\begin{center}
\includegraphics[height=6.0cm]{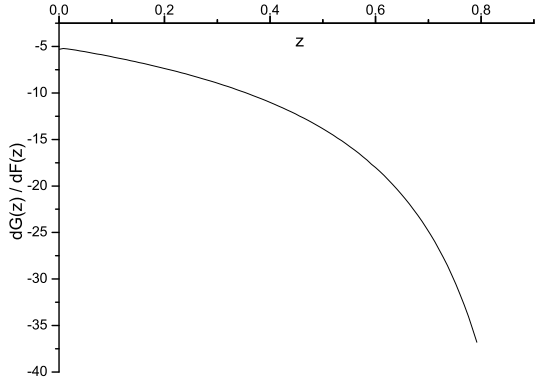}
\caption{Distribution of the scaled variable $z$ for the invariant
mass of the lepton pair. We use $F, G$ to signify the short-distance
coefficients $F({}^3S_1)$ and $G({}^3S_1)$.} \label{fig-gamma-dif}
\end{center}
\end{figure}

Employing (\ref{short-diff}), we are able to provide the following discussions.
It is instructive to look at the ratio
\begin{eqnarray}
\label{ratio-t}
t(z)\equiv \frac{d\,G({}^3S_1)}{dz}/
\frac{d\,F({}^3S_1)}{dz}=\frac{f_2(z)}{f_0(z)},
\end{eqnarray}
which solely depends on variable $z$. This
ratio characterizes the importance of the NLO
relativistic corrections compared to the LO
contribution. To visualize the relation, we plot the
ratio $t(z)$ over the variable $z$ in Fig.~\ref{fig-gamma-dif}. 
From this figure, we see that the ratio $t(z)$ is negative in the
physical region with the variable $z$ ranging from 0 to 1.
We also notice that
the magnitude of $t(z)$ rises rapidly with the increase of $z$.

In addition,
we go further to analyze the two limits of the ratio $t(z)$.
In the limit of $z\to 0$, there is
\begin{eqnarray}
\label{ratio-t-lim-1}
\lim_{z\to 0} t(z)& = &
 \frac{132-19 \pi^2}{12(\pi^2 -9)}= -5.32
 \,,
\end{eqnarray}
which agrees with the ratio of the short-distance
coefficient of the NLO relativistic corrections and that of the LO
for the processes
$\psi \to \gamma gg$ and $\psi
\to g gg$, as expected.
In the limit of $z\to 1$, it follows from
Eqs.~(\ref{eq:decayrate-result0}) and (\ref{eq:decayrate-result2}) that
$f_0(z) \to 0$, and $f_2(z) \to {\rm const}$. As a consequence,
\begin{eqnarray}
\lim_{z\to 1} t(z) & = & -\,\frac{8}{ 1-z} +\frac{4}{3} +{\mathcal O}(1-z) \,.
\label{limit-z-1}
\end{eqnarray}
From (\ref{limit-z-1}),
we see that the ratio $t(z)$ goes to infinity in the limit of $z\to 1$,
which is the result of a vanishing $f_0(z)$ in that limit.
In fact, we can see that $f_0(z)$ vanishes in the limit of $z\to 1$ from amplitude.
When the momenta of two real gluons are soft, the amplitude of $J/\psi\to \gamma^*gg$
can be separated into
\begin{eqnarray}\label{z-1}
{\mathcal A}(J/\psi\to \gamma^*gg)&=&g_s^2\bigg(\frac{p_1\cdot\epsilon_1
p_1\cdot \epsilon_2}{p_1\cdot k_1 p_1\cdot k_2}+\frac{p_2\cdot\epsilon_1
p_2\cdot \epsilon_2}{p_2\cdot k_1 p_2\cdot k_2}
-\frac{p_1\cdot\epsilon_1
p_2\cdot \epsilon_2+p_2\cdot\epsilon_1
p_1\cdot \epsilon_2}{p_1\cdot k_1 p_2\cdot k_2}
\bigg)\nonumber\\
&&\times \frac{\delta^{a_1a_2}}{2}{\mathcal A}(J/\psi\to \gamma^*),
\end{eqnarray}
where $\epsilon_{i}$ and $a_i$ indicate the polarization vector and color
index of the $i$ gluon. At LO in $v$, there is $p_1=p_2=\tfrac{P}{2}$,
and therefore ${\mathcal A}(J/\psi\to \gamma^*gg)$ vanishes. Consequently,
$f_0(z)$ vanishes in $z\to 1$.

Figure~\ref{fig-gamma-dif} and Eq.~(\ref{ratio-t-lim-1}) combine to
indicate
that the NLO relativistic corrections in this process are
not only large but increase rapidly with the rise of the virtuality of
the intermediate photon. One may doubt the convergence of the expansion
series in $v$. In Ref.~\cite{Bodwin:2002hg}, the authors calculated
the relativistic corrections to the decay rate of $\Upsilon\to ggg$ up to $v^4$.
Their results indicate the relativistic corrections from the color-singlet
matrix elements are convergent. Since the Feynman graphs are quite similar,
we expect the relativistic expansion will be convergent in the process of
$\Upsilon\to l^+l^-gg$.

The short-distance coefficients $G({}^3S_1)$ and $F({}^3S_1)$ can
be readily obtained  by integrating out the variable $z$.
Finally, substituting the short-distance coefficients given in
Eqs.~(\ref{short-diff}) into Eq.~(\ref{factor}),  we present the
differential decay rates in the NRQCD factorization formula for the
process $H(^3S_1)\to l^+l^-gg$:

\begin{eqnarray}
\frac{d\Gamma[H(^3S_1)\to l^+l^-gg]}{ dz}&=& \bigg(\frac{d
F({}^3S_1)}{dz}+ \frac{d
G({}^3S_1)}{dz}\langle {v^2}\rangle_{H}\bigg)
\frac{\langle {\mathcal O} \rangle_{H}}{m^2},
\label{Gamma-diff-final}
\end{eqnarray}
where the matrix element $\langle {v^2}\rangle_{H}$ is previously
defined in (\ref{me-ratios}). The decay rate is correspondingly
achieved by
integrating out the variable $z$.

The differential short-distance coefficients as well as the decay
rate for the lepton pair production can be easily
extended to the process $\Upsilon(nS)\to
c\bar{c}gg$, where the charm pair is produced through one virtual
gluon instead of the virtual photon. One can get them by multiplying a
color factor 5/24, and substituting $m_l$ and $e_Q^2\alpha^2$ into
$m_D$ and
$\alpha_s^2$ on the right-hand side of (\ref{short-diff}) and
(\ref{Gamma-diff-final}).\footnote{In calculating the decay rate
of the process $\Upsilon(nS)\to
c\bar{c}gg$, we take the mass of the charm quark to be that of the 
$D$ meson $m_c=m_D$ in order to compare with the measurement of the
experiment~\cite{Chung:2008yf}.}

\section{Numerical Results and discussions}
\label{sec:conclusion}

In this section, we first numerically evaluate the total decay rate and the
short-distance coefficients for various processes, and then discuss the
momentum distribution related to the charmed meson $D^{*+}$ in the process
$\Upsilon(1S)\to c\bar{c}gg\to D^{*+}X$.

\subsection{Decay rate and the short-distance coefficients}

In this subsection, we employ the obtained differential short-distance
coefficients (\ref{short-diff}) and the decay rate
(\ref{Gamma-diff-final})
to make numerical predictions for the decay rate of the processes
$H(^3S_1)\to l^+l^-(c\bar{c})gg$. The corresponding discussions are also
presented.

\begin{table}
\centering
\caption{\label{table1} Numerical values for the parameters of
different initial-state particles: the mass $m_H$, strong coupling
constant $\alpha_s(m_H/2)$,
the NRQCD matrix elements  $\langle {\mathcal O}_1\rangle_H$ 
and the value of $\langle v^2 \rangle_{H}$.}
\begin{tabular}{lccccccccc}
\hline
& &$m_H(\rm GeV)$&  &$\alpha_s (m_H/2)$&  &$\langle{\mathcal O}_1\rangle_{H}(\rm GeV^3)$& &$\langle v^2 \rangle_{H}$&\\
\hline
$J/\psi$ & &3.097&&0.334& & 0.440& & 0.225&\\
\hline
$\psi (2S)$ & &3.686& &0.300&  & 0.274& &0.633 &\\
\hline
$\Upsilon (1S)$ & &9.460& &0.215& & 3.07& &0.057 &\\
\hline
$\Upsilon (2S)$ & &10.023& &0.211&  & 1.62& &0.179 &\\
\hline
$\Upsilon (3S)$ & &10.355& &0.210&  & 1.28& &0.251 &\\
\hline
\end{tabular}
\end{table}

\begin{table}
\centering
\caption{\label{table2} The ratio $r$, theoretical
predictions for the decay rate, the ratio
between the NLO rate and LO rate, and the ratio between the short-distance
coefficients.}
\begin{tabular}{lccccccccccc}
\hline
&&$r$ &&$\Gamma^{(0)}(\rm keV)$&  &$\Gamma^{(2)}(\rm keV)$& & $\Gamma^{(2)}/\Gamma^{(0)}$& &$G_{1}({}^3S_1)/{F_{1}({}^3S_1)}$\\
\hline
$J/\psi \to e^+ e^-gg$ &&$1.08\times10^{-7}$ &&$4.73 \times 10^{-1}$&&$-5.91\times 10^{-1}$& &$-125\%$&& -5.56&\\
\hline
$J/\psi \to \mu^+ \mu^-gg$ &&$4.69\times10^{-3}$&&$1.08 \times 10^{-1}$& &$-1.57\times 10^{-1}$& &$-145\%$& & -6.49&\\
\hline
$\psi (2S) \to e^+ e^-gg$ && $7.66\times10^{-8}$&&$2.43 \times 10^{-1}$& &$-8.56\times 10^{-1}$& &$-352\%$& & -5.55&\\
\hline
$\psi (2S) \to \mu^+ \mu^-gg$ &&$3.31\times10^{-3}$ &&$5.98 \times 10^{-2}$& &$-2.41\times 10^{-1}$& &$-403\%$& & -6.37&\\
\hline
$\Upsilon (1S) \to e^+ e^-gg$ &&$1.16\times10^{-8}$&&$3.68 \times 10^{-2}$& &$-1.16\times 10^{-2}$& &$-31.5\%$& & -5.53&\\
\hline
$\Upsilon (1S) \to \mu^+ \mu^-gg$ &&$5.02\times10^{-4}$ &&$1.22 \times 10^{-2}$& &$-4.16\times 10^{-3}$& &$-34.0\%$& & -5.97&\\
\hline
$\Upsilon (1S) \to \tau^+ \tau^-gg$ &&$1.41\times10^{-1}$&&$1.05 \times 10^{-3}$& &$-7.06\times 10^{-4}$& &$-67.3\%$& & -11.8&\\
\hline
$\Upsilon (1S) \to  c\bar c gg$ &&$1.56\times10^{-1}$ &&$1.44$& &$-1.01$& &$-70.4\%$&
& -12.4&\\
\hline
$\Upsilon (2S) \to c\bar c gg$ &&$1.39\times10^{-1}$ &&$7.99 \times 10^{-1}$& &$-1.68$&
&$-210\%$& & -11.7&\\
\hline
$\Upsilon (3S) \to c\bar c gg$ &&$1.30\times10^{-1}$ &&$6.63 \times 10^{-1}$& &$-1.90$&
&$-287\%$& & -11.4&\\
\hline
\end{tabular}
\end{table}
To this end, we need to
specify various input parameters, such as
the coupling constants, the pole masses of the heavy quarks,
the physical masses of various
involved quarkonia and final-state leptons and charm quark
(we choose the mass of the final-state charm quark
to be the mass of the charmed hadron),
and the values of the
nonperturbative NRQCD matrix elements.
In our calculation, we take the charm and bottom quark pole masses to be
$m_c=1.4$ GeV and $m_b=4.6$ GeV, respectively.
The lepton masses are taken to be $m_e = 0.51\times 10^{-3}$ GeV,
$m_{\mu}=0.106$ GeV, $m_{\tau}=1.777$ GeV~\cite{Nakamura:2010zzi}.
Since the final-state charm quark will dominantly evolve to the charmed hadron,
we choose
the charm quark mass to be the mass of the charmed hadron $m_D=1.87$ GeV,
which is the average masses of the $D^0$ and $D^+$.
The fine structure constant changes slightly from the scale of charmonium to that
of bottomonium, so we uniformly choose $\alpha=\frac{1}{133}$
for all the decay processes involved.

The values of the quarkonium masses, coupling constants, 
and the NRQCD matrix elements are listed in Table~\ref{table1},
where scales of the
coupling constants are chosen to be
half of the corresponding decay quarkonium. In the table, the masses
of the quarkonia are taken from Ref.~\cite{Nakamura:2010zzi};
we take the NRQCD matrix elements $\langle {\mathcal O}\rangle_{J/\psi}$
and $\langle {v^2}\rangle_{J/\psi}$
from
Ref.~\cite{Bodwin:2007fz},  $\langle {\mathcal O}\rangle_{\Upsilon(nS)}$ from
Ref.~\cite{Kang:2007uv}, and
$\langle {\mathcal O}\rangle_{\psi(2S)}$ from Ref.~\cite{HJCH:2008};
other values of the NRQCD matrix elements
$\langle {v^2}\rangle_{H}$ are
determined by the Gremm-Kapustin relation~\cite{Gremm:1997dq}:
\footnote{Since the pole masses of the charm quark and bottom quark
are not determined very well,
the NRQCD matrix element computed from
the Gremm-Kapustin relation has a large uncertainty. This is especially serious for
the bound state quarkonium, whose mass is close to $2m_{pole}$.
Therefore, in the next subsection, we adopt a new method to determine
$\langle v^2\rangle_H$ for $\Upsilon(1S)$.}
\begin{eqnarray}
\langle v^2\rangle_H=\frac{m_H-2m_{pole}}{m_{pole}},
\end{eqnarray}
where $m_{pole}$ denotes the pole mass of the heavy quark, which is
taken to be $1.4$ GeV and $4.6$ GeV for the charm quark and bottom quark, 
respectively.

With the parameters chosen above, we are able to make numerical
predictions for various
decay channels, which include the inclusive lepton decay of the
charmonium and bottomonium,
as well as the inclusive charm decay of the
bottomonium.
First, we consider the total decay rate. The predicted results are listed in
Table~\ref{table2}.  In the table,
we give the decay rates both in the LO and
in the NLO relativistic corrections. To show the magnitude of the
relativistic corrections, we also list two ratios. One is the ratio
of the NLO and the LO short-distance coefficients, namely,
$G(^3S_1)/F(^3S_1)$. The other is the ratio of
the NLO and the LO decay rates $\Gamma^{(2)}/\Gamma^{(0)}$.

From Table~\ref{table2},
we find that all
the relativistic corrections are huge and negative. This is
especially serious for the bottomonium decay to charm pair channels.
We can reach two conclusions from the table. First, the
ratio of the NLO and the LO short-distance coefficients
ascends with the increase of $r$, which is previously defined  as
$4m_{l}^2/m_H^2$ [or $4m_{D}^2/m_H^2$ for
$\Upsilon(nS)\to c \bar c gg$]. Second, in the channel with small $r$
such as $\Upsilon(nS)(\psi(nS))\to e^+e^-gg$, the ratio
of the short-distance
coefficients approaches to that in the process of $J/\psi \to \gamma gg$ or
$J/\psi \to ggg$. This is
understood from the fact that the decay rate of $H(^3S_1)\to e^+e^-gg$
is dominated by the region,
where the virtual photon is nearly on-shell.

It is also intriguing to study the $r$ dependence of the
relativistic corrections.
\begin{figure}[tb]
\begin{center}
\includegraphics[height=6.0cm]{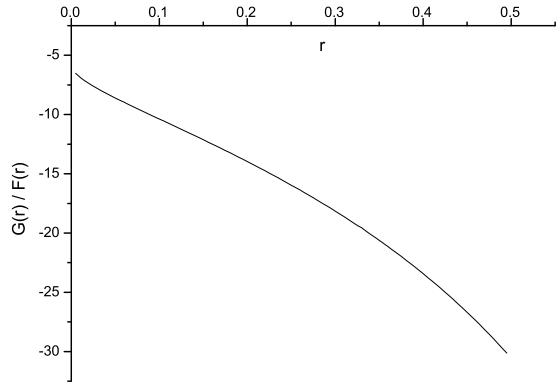}
\caption{Dependence of the ratio of the short-distance coefficients $G(^3S_1)/F(^3S_1)$
on $r$.}
\label{fig-gamma-r}
\end{center}
\end{figure}
In Fig.~\ref{fig-gamma-r}, we show the dependence of
the ratio $G(^3S_1)/F(^3S_1)$ on $r$.
From the figure, we see that as the value of $r$ increases, the
relativistic corrections will
increase rapidly. Actually, this feature has been shown in
Table~{\ref{table2}}.
When the mass of the final-state fermion is close to half of that
of the initial quarkonium,
the momenta of the two real gluons will become soft, and therefore
the perturbative QCD calculation is
unreliable. Therefore,
only the region $r<0.5$ is plotted in Fig.~\ref{fig-gamma-r}.

\subsection{Momentum distribution of charmed hadron $D^{*+}$}

To predict the production
rate of a charmed hadron from $\Upsilon(1S)$ decay, we need to
consider the
probability of a charm quark hadronizing into the charmed hadron. In Ref.~\cite{Seuster:2005tr},
the authors computed the ratio ${\rm Br}[c\to h]$.
In the Table~10 of Ref.~\cite{Seuster:2005tr},
one can read that the ratio for $D^{*+}$ production is
${\rm Br}[c\to D^{*+}]=0.220$. With this value, we can readily derive the decay rate
for $D^{*+}$ production through the process
$\Upsilon(1S)\to c\bar{c}gg\to D^{*+}X$.

As mentioned in the previous subsection, the NRQCD matrix element
$\langle v^2\rangle_{\Upsilon(1S)}$ determined from the Gremm-Kapustin relation
is sensitive to the bottom pole mass. Here we present another method to determine
this matrix element, and then use the new value to
predict the momentum distribution of $D^{*+}$.

In Ref.~\cite{:2009wm}, the \textsl{BABAR} Collaboration reported their measurement
${\rm Br}\left[\Upsilon(1S) \to D^{*+}X\right]=(2.52\pm0.13
({\rm stat})\pm0.15({\rm syst}))\%$.
In addition, they derived the contribution from the virtual photon
annihilation process to be
${\rm Br}\left[\Upsilon(1S) \to \gamma^* \to D^{*+}X\right]=(1.52\pm0.20)\%$,
and therefore we may expect that the difference arises
from the contribution of $\Upsilon(1S) \to c\bar{c}gg \to D^{*+}X$.
With this assumption,\footnote{In Ref.~\cite{Zhang:2008pr}, the authors
considered the contribution to the charm pair production
from the color-octet NRQCD matrix element. According to the NRQCD
velocity-scaling
rules,
this contribution belongs to the higher order corrections at $v$ expansion.
We are now working on $v^4$ corrections to
the process $\Upsilon\to c\bar{c}gg$, and
a thorough analysis including the contributions from the 
color-octet NRQCD matrix elements
will be presented in the future.}
we are able to fix the
value of $\langle v^2\rangle_{\Upsilon(1S)}$
through the
relation\footnote{Since we use the experimental data
related to
the $D^{*+}$ production, here we choose $r=4m^2_{D^{*+}}/
m^2_{\Upsilon(1S)}$, where $m_{D^{*+}}=2.01$ GeV.}
\begin{eqnarray}
\frac{1}{\Gamma_\Upsilon}
\bigg(F(^3S_1)+ G(^3S_1)\langle v^2\rangle_H\bigg)
\times \frac{\langle{\mathcal O}_1\rangle_H}{m_b^2}\times {\rm Br}[c\to D^{*+}]=2.52\%-1.52\%=1.00\%,
\end{eqnarray}
where $\Gamma_\Upsilon$
represents the total decay rate of
$\Upsilon(1S)$. By taking as $\Gamma_{\Upsilon}=54.02$ keV,
we obtain $\langle v^2\rangle_{\Upsilon(1S)}=-0.0781$. In Ref.~\cite{Chung:2010vz},
this matrix element is also determined to be $-0.009\pm0.003$ through fitting
the decay rate of the process $\Upsilon\to e^+e^-$. Though both results are
negative, our result is much larger than theirs.
We apply the formulas (\ref{dif-integration-1}) and
(\ref{dGamD}) derived in
Sec.~\ref{sec:dif-L} to calculate the momentum distribution of $D^{*+}$.
Prior to making the numerical predictions, we need to select an 
appropriate fragmentation function. Here
we employ two well-known models:
the Kartvelishvili-Likhoded-Petrov (KLP)
fragmentation function~\cite{Kartvelishvili:1977pi}, which was used
in the analyses of charmed-hadron momentum distribution in $\Upsilon(nS)$
and $\chi_b$
decays, and the Peterson fragmentation function~\cite{Peterson}.
The KLP and Peterson fragmentation functions both have a simple
parametrization depending only on the light-cone momentum fraction
$z^\prime$ (see Table~\ref{klp-p}).
The optimal values of $\alpha_c$ determined by the Belle
Collaboration are $\alpha_c=5.6$, and $0.054$ for the KLP  and
Peterson fragmentation functions, respectively~\cite{Seuster:2005tr}.

The normalization factor $N_h$ is determined by the constraint
$\int_0^1 dz D_{c\to h}(z)=$Br$[c\to h]$.
Taking the fragmentation probability ${\rm Br}[c\to D^{*+}]=0.220$,
we are able to determine the normalization factors for the two
fragmentation functions,
which are shown in Table~\ref{klp-p}.
\begin{table}
\centering
\caption{\label{klp-p} The KLP and Peterson fragmentation function and the value of the corresponding parameters.}
\begin{tabular}{lc|cccccccccc}
\hline
&    & & $D(z^\prime)$& & $N_h$& &$\alpha_c$&\\
\hline
&KLP&  & $N_h z^{^\prime\alpha_c(1-z^\prime)}$& & 11.0& &5.6&\\
\hline
&Peterson& & $N_h \frac{1}{z^\prime}(1-\frac{1}{z^\prime}
-\frac{\alpha_c}{1-z^\prime})^{-2}$& &0.127& &0.054&\\
\hline
\end{tabular}
\end{table}

\begin{figure}[ht]
\begin{center}
\includegraphics*[scale=0.6]{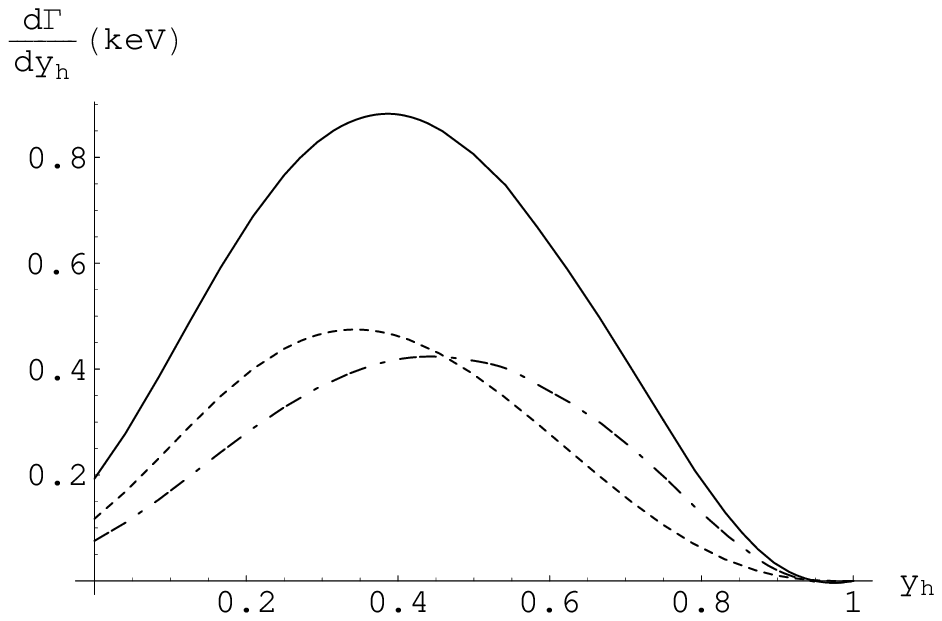}
\includegraphics*[scale=0.6]{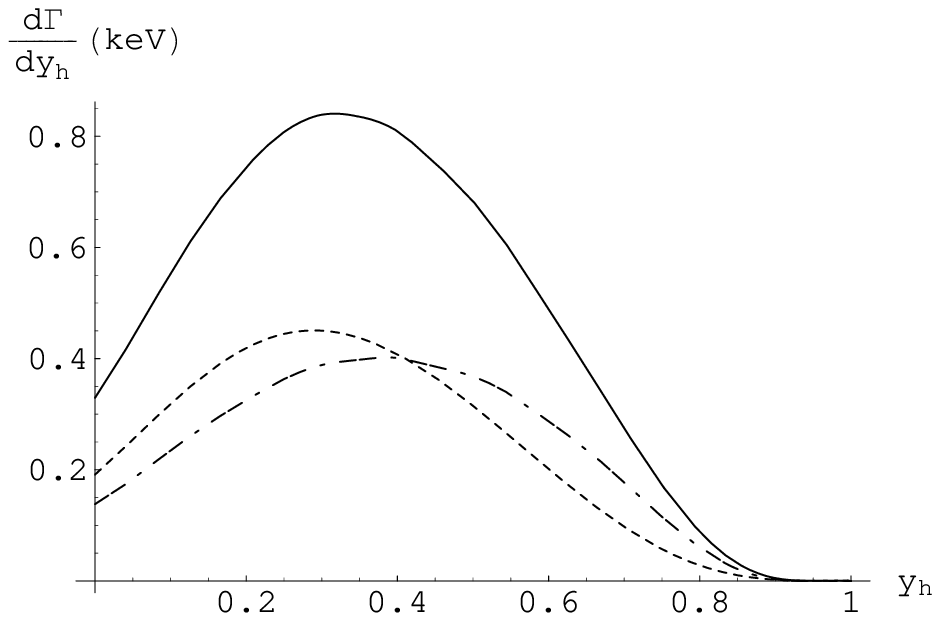}
\caption{The momentum distribution of the charmed hadron $D^{*+}$ for the different fragmentation function. The left figure is for the KLP model, and the right one is for the Peterson model.
In the figure, the dotted line,
dash-dotted line, and solid line correspond to the LO, NLO, and total
distributions, respectively.}
\label{fig-distribution}
\end{center}
\end{figure}

With the formulas (\ref{dif-integration-1}) (\ref{dGamD}) and the
fragmentation functions in
Table~\ref{klp-p}, we can evaluate the momentum distribution of $D^{*+}$
in the KLP and Peterson models, which
is shown in Fig.~\ref{fig-distribution}.
We notice that the discrepancy between the figures
from the two models is small. This implies
the momentum distribution is insensitive to the models. In addition,
we find that the contribution from the NLO
relativistic corrections is comparable with
that of the LO, and therefore modifies the LO magnitude significantly.

\section{Summary}
\label{sec:summary}

In this work, we compute the NLO relativistic corrections to the decay rates of
the processes
of $\psi(nS)(\Upsilon(nS))\to l^+l^-(c\bar{c})gg$ in the framework of the NRQCD factorization formula.
The differential short-distance coefficients and decay rates over the
invariant mass of the lepton pair or the charm
pair are presented analytically.
The relativistic corrections to all the processes
are significant. The magnitude of the NLO relativistic corrections
even surpasses that of the LO contribution in most processes.
Furthermore, we analyze the ratio of the
differential short-distance coefficients. The results indicate that
the relativistic corrections increase rapidly with rise of
the invariant mass of the lepton pair
or the charm pair. In addition, we study the $r$ dependence of the ratio of
the short-distance coefficients $G(^3S_1)/F(^3S_1)$. In the limit of $r\to 0$,
our result is consistent with
that of $J/\psi\to \gamma gg$ or $J/\psi\to ggg$. With the increase of $r$, the ratio $G(^3S_1)/F(^3S_1)$
increases rapidly.

The momentum distributions of a free charm quark and of a charmed hadron in the process
$\Upsilon(1S)\to c\bar{c}gg\to D X$ are studied.
We also
determine the NRQCD matrix element $\langle v^2\rangle_{\Upsilon(1S)}$
based on the measurement of the \textsl{BABAR} Collaboration.
Taking it as an input parameter, we also predict the momentum distribution of
$D^{*+}$
through the process $\Upsilon(1S)\to c\bar{c}gg\to D^{*+}X$.

\begin{acknowledgments}
We thank Bin Gong for helpful discussions.
The research of H. C. and Y. C. was supported by the NSFC with Contract No. 10875156.
The research of W. S.
was supported by the National Natural Science Foundation of China
under Grants No. 10875130 and No. 10935012 and  by the Basic
Science Research Program through the NRF of Korea funded by the MEST under Contract
No. 2011-0003023.
\end{acknowledgments}


\end{document}